\documentclass{article}
\usepackage{microtype}
\usepackage{graphicx}
\usepackage{booktabs}
\usepackage{multirow}
\usepackage{makecell}
\usepackage{colortbl}
\usepackage{xcolor}
\usepackage{multicol}     
\usepackage{array}  
\usepackage{amsmath}
\usepackage{enumerate}
\usepackage{subcaption}
\usepackage{hyperref}
\usepackage{cleveref}
\usepackage{xurl}
\usepackage[shortlabels]{enumitem}
\usepackage[moderate,leading=normal,tracking=normal,wordspacing=normal,lists=normal]{savetrees}
\usepackage{float}
\usepackage{stfloats} 

\usepackage[accepted]{mlsys2024}
\definecolor{codecolor}{RGB}{70,70,180} 
\definecolor{registercolor}{RGB}{160,40,40} 
\definecolor{immediatecolor}{RGB}{0,110,80}  
\definecolor{memorycolor}{RGB}{100,50,140}
\definecolor{labelcolor}{RGB}{180,90,40}
\newcommand{\code}[1]{{\color{codecolor}\textbf{\texttt{#1}}}}
\newcommand{\reg}[1]{{\color{registercolor}\textbf{\texttt{#1}}}}
\newcommand{\imm}[1]{{\color{immediatecolor}\textbf{\texttt{#1}}}}
\newcommand{\mem}[1]{{\color{memorycolor}\textbf{\texttt{#1}}}}

\newcommand{\titlename}{From CISC to RISC: \\Language-Model Guided Assembly Transpilation}
\newcommand{\titlenameRunning}{From CISC to RISC: Language-Model Guided Assembly Transpilation}

\mlsystitlerunning{\titlenameRunning}

\begin{document}
\twocolumn[
\mlsystitle{\titlename}
\begin{center}
{\large
\mlsyssetsymbol{equal}{\normalsize *}
\begin{mlsysauthorlist}
\mlsysauthor{Ahmed Heakl}{equal}
\mlsysauthor{Chaimaa Abi}{equal}
\mlsysauthor{Rania Hossam}{}
\mlsysauthor{Abdulrahman Mahmoud}{}
\end{mlsysauthorlist}
\vspace{2ex}
{\normalsize Mohamed bin Zayed University of Artificial Intelligence, Abu Dhabi, UAE}
}
\end{center}
\mlsyskeywords{Machine Learning, MLSys}
\vskip 0.3in
\begin{abstract}
The transition from x86 to ARM architecture is becoming increasingly common across various domains, primarily driven by ARM's energy efficiency and improved performance across traditional sectors. However, this ISA shift poses significant challenges, mainly due to the extensive legacy ecosystem of x86 software, and lack of portability across proprietary ecosystems and software stacks. This paper introduces CRT, a lightweight LLM-based transpiler that automatically converts x86 assembly to ARM assembly. Our approach bridges the fundamental architectural gap between x86's CISC-based and ARM's RISC-based computing paradigms while preserving program semantics and optimizing performance.
We evaluate CRT on diverse real-world applications, achieving 79.25\% translation accuracy from x86 to ARMv5 on our comprehensive test suite, and a 88.68\% accuracy from x86 to RISC-V. In practical deployments on Apple M2 hardware (ARMv8), our transpiled code achieves 1.73x speedup compared to Apple's Rosetta 2 virtualization engine, while delivering 2.41x memory efficiency and 1.47x better energy consumption.  Through testing and analysis, we show that CRT successfully navigates the CISC/RISC divide, and generates correctly executable RISC code despite machine ''language'' barriers. We release our code, models, training datasets, and benchmarks at: \url{https://ahmedheakl.github.io/asm2asm/}


\end{abstract}]
\equalcontributionnotice
\section{Introduction}
\label{intro}

The ending of Moore's Law and Dennard scaling has led to a paradigm shift in the way modern processors are designed and architected. No longer benefiting from generational improvement in power, performance, and area efficiency (PPA), many academic and industry players are rethinking their architectural designs. This includes both introducing more specialized hardware components (such as tensor cores for ML processing)~\cite{markidis2018nvidia,jouppi_et_al_2017}, alternatives to Von Neumann architectures (e.g., processing-in-memory~\cite{han_et_al_2020,kang_et_al_2013}, and revisting the fundemental computing paradigm of CISC versus RISC designs.

Complex instruction set computers (or CISC) such as Intel and AMD's x86 instruction set architecture (ISA) have maintained a stronghold in the datacenter and server space (holding more than 79.9\% of the market~\cite{arm-servers} as well as personal computing devices (with more than 82\% market share in 2024~\cite{arm-pc}). Reduced instruction set computers (RISC), on the other hand, are predominantly deployed in energy-constrained environments, such as IoT devices, mobile phones, and edge devices~\cite{arm-servers}. In this space, ARM corporation is the predominant player, licensing its architecture to many companies such as Apple and Qualcomm mobile phone chips, and ``smart" devices like household appliances. However, this has all begun changing very fast in recent years~\cite{woo_2020_rise_arm}.

For example, Apple recently switched over from Intel x86 chips to power their Macbooks and are now exclusively using ARM chips rebranded as M1, M2, and M3 cores~\cite{shilov_2023}. Amazon has begun designing in-house chips called Graviton~\cite{nextplatform_graviton3} to power their datacenter and compute services. Furthermore, Microsoft has also been adopting their operating system design to support ARM, both for in-house hardware (such as Surface tablets) as well as general purpose compute (for Windows OS). While many business decisions go into such a drastic change in infrastructure, the common thread is that ARM-based devices provide excellent energy-efficiency compared to their x86 counterparts, and have dramatically bridged the gap in performance as well~\cite{cloudpanel_arm_servers}.

Despite lots of potential advantages of low-powered ARM processors, adopting or moving to a new ISA in the hardware space renders prior code in the software space incompatible, as the machine code must be re-targeted for the new ISA. The ISA defines the hardware-software ``contract", which has enabled substantial improvements over the years independently in the software space and hardware space. Nevertheless,
an alternative ``machine language" presents a huge technological barrier to overcome, as code portability now becomes a problem. Additionally, for competitive reasons, many companies do not ship their source code around, and instead provide executable binaries which are challenging to de-compile and break intellectual property (IP).

Alternatively, to avoid having to recompile a huge code base and to address code portability challenges, hardware virtualization has been used to dynamically ``translate" between ISAs. 
Several tools currently facilitate x86 to ARM virtualization. QEMU~\cite{qemu}, an open-source, general-purpose emulator supporting multiple architectures and, while versatile, introduces significant performance overhead compared to native execution~\cite{wei2019performance}. Rosetta 2~\cite{rosetta2}, Apple's proprietary translation layer, specifically designed for x86 to ARM (Apple Silicon) translation, is more efficient than QEMU, but its closed-source nature and platform-specificity limit its broader application. Given these existing solutions' limitations, a crucial question emerges: Can we develop a direct translation medium between x86 and ARM architectures that ensures correctness without the performance hit of a virtualization layer?

Addressing the \textit{correctness} of assembly to assembly translations is arguably the more challenging research aspect. 

In addition to legacy software compatibility reasons, CISC instructions have fundamental architectural differences to RISC instructions, such as how an instruction handles registers versus memory for different opcodes, the number of available general purpose registers between the ISAs, and the access of sub-register bits (e.g., modifying only the lower 8-bits of a register value). Similarly, compilers need to work harder to generate ARM-based code, while that complexity is lowered onto the micro-architecture in an x86-based system. Binary sizes are also different, where programs assembled into x86 are typically much shorter (due to the complex nature of instructions which are later converted to micro-code inside the processor), compared to the longer binaries in ARM (composed of many, simpler instructions). Furthermore, operating systems need to be aware of the underlying ``language" of the hardware in order to properly manage it; we observe this phenomenon when seeing that Windows today has two separate OS binaries for x86 versus ARM~\cite{anderson2023windows,microsoft2024windowsarm}.

Existing approaches demonstrate an apparent dichotomy: open-source emulators prioritize flexibility over performance, while solutions like Rosetta achieve efficiency through hardware-specific optimizations. This creates a gap in the ecosystem for a solution that combines both attributes.

This paper proposes an approach based on language models (LMs) as machine translation engines. Building upon LMs' demonstrated success in various translation tasks, we apply them to learn mappings between x86 and ARM assembly code through paired examples. Intuitively, we use the LM to ``translate" between CISC and RISC machine code, just as it has performed admirably in recent years in enabling translation between human languages. A key difference, however, is that the translation must be \textit{precise} in machine languages, as any incorrect syntax, mis-used registers, bad jumps, or other architecturally important constructs can render the program incorrect semantically and/or functionally.
Our methodology presents an opportunity to merge open development practices with high-performance translation capabilities, as the models can identify and optimize instruction patterns without depending on proprietary optimizations or predetermined translation rules while avoiding complete system emulation overhead or code overhauls.

This work presents the following key contributions:
\begin{enumerate}
    \item The first CISC to RISC transpiler, coined CRT, built via a custom-trained LLM achieving a test accuracy of 79.25\% on ARM and 88.69\% on RISC-V64.  
    \item An in-depth analysis into the inner workings of our transpiler, including hardware-informed design decisions to best train an accurate LLM model for assembly transpilation (\S\ref{approach}, \S\ref{discussion}).
    \item We perform a case-study using our transpiler in a real-world setting, and compare it to Apple Rosetta's x86 to ARM virtualization engine. Results show that CRT's generated assembly achieves 1.73x speedup compared to Rosetta while delivering 1.47x better energy efficiency and 2.41x memory efficiency (\S\ref{study}). 
\end{enumerate}

In the remainder of this paper, we provide additional background in the space of assembly languages and LLM-for-hardware (\S\ref{background}), describe our approach in designing the transpiler (\S\ref{approach}), evaluation of its efficacy (\S\ref{eval}), and discuss the challenges and benefits of our approach (\S\ref{sec_results}, \S\ref{discussion}). Finally, we conclude with a case study of transpiling for an Apple M2 processor, and the advantages over prior approaches for this task (\S\ref{study}).

\section{Background \& Related Work}
\label{background}

\textbf{Virtualization and Emulation:} Emulation and assembly-level virtualization enable the execution of one ISA's binary on a host machine for which it was not compiled for originally. QEMU \cite{bellard2005qemu}, an open-source emulator, uses dynamic binary translation \cite{sites1993binary}, offering flexibility but with performance overhead, enabling x86 to ARM emulation, amongst other ISAs. Rosetta 2, Apple's virtualization layer for macOS, combines ahead-of-time (AOT) and just-in-time (JIT) translation, providing better performance within the Apple ecosystem. 

These approaches face challenges in achieving native-level performance and ensuring broad compatibility, due to the dynamic nature of execution. A transpiler approach, directly converting x86 to ARM assembly, could supplant these solutions by eliminating runtime translation overhead with a one-time translation into the host ISA. This method could address the limitations of current emulation and virtualization techniques, particularly in performance-critical scenarios, or where pre-processing is feasible, or when source code is not available (due to proprietary IP). 

\textbf{Neural Code Translation:} Machine learning approaches for code translation have primarily focused on high-level programming. Initial research explored neural machine translation architectures, with TransCoder \cite{lachaux2020unsupervised} demonstrating unsupervised translation between C++, Java, and Python. Pre-trained transformer models such as CodeBERT \cite{feng2020codebert} and CodeT5 \cite{wang2021codet5} have advanced code understanding and generation capabilities across multiple programming languages. The emergence of LLMs specialized for code, including Code Llama \cite{roziere2023code} and DeepSeek Coder \cite{liu2024deepseek}, has demonstrated increasingly sophisticated code manipulation capabilities through self-supervised learning on vast code repositories. These methods often rely on structural representations like abstract syntax trees for high-level language translation \cite{chen2018tree}. However, these methods face distinct challenges when applied to assembly-level translation, where the structural representations and semantic preservation requirements differ significantly from high-level language translation.

\textbf{Language Models for Low-Level Programming:} Recent research has increasingly demonstrated the potential of language models in various tasks related to low-level code analysis and transformation. A language model is a statistical model that learns to predict the probability distribution of tokens within a language, enabling it to generate coherent and contextually relevant text. 
These models have been successfully applied in areas such as decompilation, binary similarity analysis, and compiler optimization, demonstrating their ability to tackle intricate, instruction-level challenges.

In decompilation, LLM4Decompile \cite{tan2024llm4decompile} introduced specialized language models for direct binary-to-source translation and decompiler output refinement. DeGPT \cite{hu2024degpt} further explored decompiler enhancement through semantic-preserving transformations. SLaDe \cite{armengol2024slade} combines a 200M-parameter sequence-to-sequence Transformer with type inference techniques to create a hybrid decompiler capable of translating both x86 and ARM assembly code into readable and accurate C code, effectively handling various optimization levels (\verb|-O0| and \verb|-O3|).

Language models have also been adapted to optimization tasks, with LLM Compiler \cite{cummins2024meta} introducing a foundation model that supports zero-shot optimization flag prediction, bidirectional assembly-IR translation, and compiler behavior emulation. This approach demonstrates the potential for language models to enhance compiler optimization workflows by automating complex tasks.

Binary similarity analysis has similarly benefited from language model adaptations. DiEmph \cite{xu2023improving} addressed compiler-induced biases in transformer models, while jTrans \cite{wang2022jtrans} incorporated control flow information into the transformer architecture. Yu et al. \cite{yu2020order} combined BERT-based semantic analysis with graph neural networks to capture both semantic and structural properties of binary code.

While these applications have shown promising results, the use of LLMs for assembly code transpilation remains relatively underexplored. Assembly languages present unique challenges due to the fundamental differences in instruction sets and execution models across architectures. GUESS \& SKETCH \cite{lee2024guess} introduced a neurosymbolic approach combining language models with symbolic reasoning for translating assembly code between ARMv8 and RISC-V architectures. Our work, CRT, extends this direction by addressing the more challenging task of transpiling between CISC (x86) and RISC (ARM, RISC-V) architectures, bridging fundamental architectural differences in ISA complexity and execution models.

\section{Approach} \label{approach}
This sections describe our approach for CISC to RISC assembly transpilation (CRT), covering the problem setup, model choices, training, and tokenizer adaptations.

\begin{figure}[!h]
    \centering
    \small
    \includegraphics[width=0.9\linewidth]{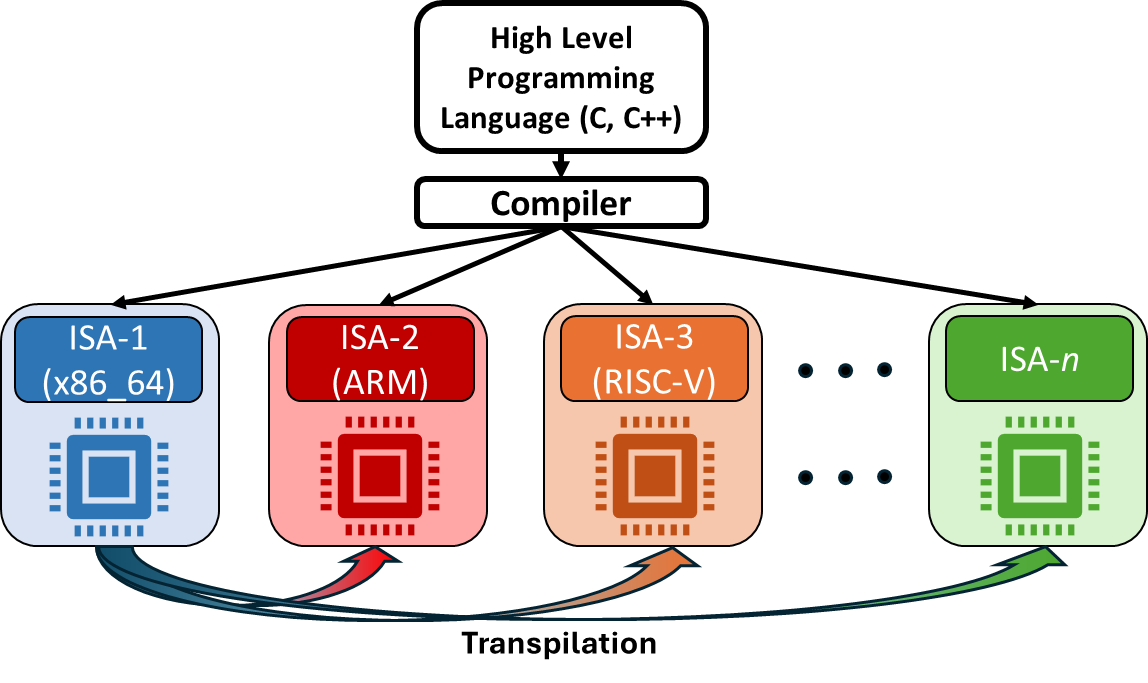}
    \vspace{-0.1in}
    \caption{Conceptual representation of an asm-to-asm transpiler, which would enable direct ``translation" from one machine language to another without needing the source code and by-passing the software stack.}
    \label{fig:transpilation}
    \vspace{-0.2in}
\end{figure}

\subsection{Problem Definition}
We aim to translate x86 assembly code to ARM assembly code by leveraging Language Models (LMs) to automatically handle the fundamental differences between these ISAs. Let \( X = \{x_1, x_2, \dots, x_n\} \) denote the set of x86 vocab, and \( Y = \{y_1, y_2, \dots, y_n\} \) denote the set of corresponding ARM vocab. Our goal is to learn a mapping function \( f: X \rightarrow Y \) that translates any x86 code \( x_i \) into its ARM equivalent \( y_i \).

\begin{figure*}[t]
    \centering
    \small
    \includegraphics[width=0.9\linewidth]{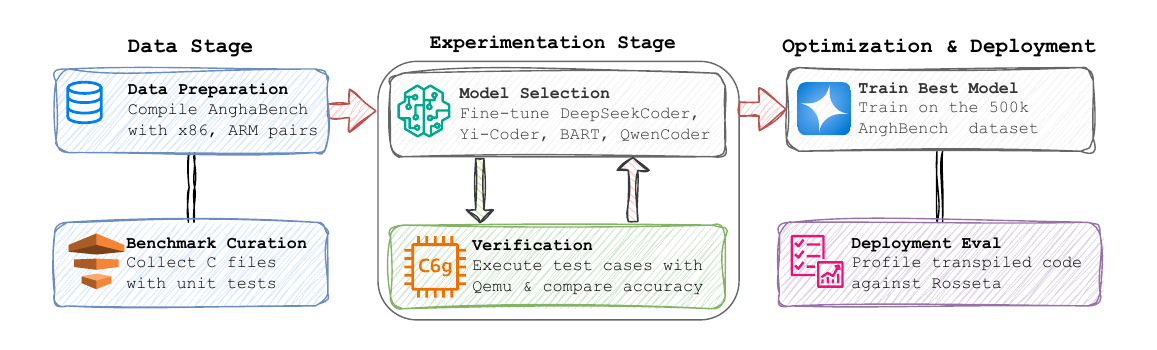}
    \vspace{-.2in}
    \caption{CRT pipeline stages: \textbf{Data} (AnghaBench data curation), \textbf{Experimentation} (model tuning and accuracy), and \textbf{Optimization \& Deployment} (final training and Rosetta evaluation).}
    \label{fig:pipeline}
    \vspace{-.2in}
\end{figure*}

Our approach is to utilize LLMs and have the model learn the conditional distribution \( P(Y \mid X; \theta) \), where \( \theta \) represents the model parameters. The model generates ARM code in an autoregressive manner, producing each token \( y_t \) based on the input x86 code \( X \) and previously generated tokens \( y_{<t} \): \( P(Y \mid X; \theta) = \prod_{t=1}^{T} P(y_t \mid y_{<t}, X; \theta) \). This approach aims to provide contextually accurate translations,  avoiding prior rule-based methods of the past~\cite{daisy1997}.

\subsection{Training Stages and Model Selection}
Our framework consists of three main stages: Data collection, Experimentation (of hyperparameters), and Optimization \& Deployment, as illustrated in Figure~\ref{fig:pipeline}.

\textbf{Data Stage:}
We construct our dataset by collecting a diverse corpus of C source files and generating corresponding x86 (CISC) and ARM (RISC) assembly pairs. These paired compilations constitute our training data, from which the language model learns the cross-architectural mapping between CISC and RISC instruction sets. For evaluation, we employ two metrics: functional correctness assessed through unit tests accompanying the C files, and assembly-level similarity measured via the more strict metric of \textit{edit distance} (\S\ref{sec:metrics}) between the generated outputs and ground truth assembly pairs.

\textbf{Experimentation Stage:} 
For model selection, we fine-tune various small-scale, established open-source language models, leveraging their efficiency in processing high-level code. During this phase, we experiment with hyperparameters including batch size, gradient accumulation~\cite{gradient-accumulation}, warmup steps~\cite{warmup}, optimizers (e.g., AdamW~\cite{adamw} and Adafactor~\cite{adafactor}), learning rates, and epochs to find the best settings for our application of assembly transpilation. Each model undergoes a thorough evaluation on our curated benchmark, and accuracy is verified by executing test cases using Qemu~\cite{qemu} to ensure reliability. We also explored the alignment of code pairs in this phase. While this showed initial promise, it did not help significantly and so we leave our experimentation details for the interested reader in appendix~\S\ref{appendix:assembly_comparison}. 

\textbf{Optimization \& Deployment:} 
In the stage, the best-performing model is selected and trained on the entire dataset. We also explore post-training quantization, to study the efficacy of compressing our asm-to-asm transpiler for a low-resource deployment setting. We run a case study to evaluate and compare the transpiled code against Rosetta~\cite{rosetta2}, to assess performance in a production-like environment (\S\ref{study}).

\subsection{Tokenizer Extension}
\begin{table}[h]
    \centering
    \small
    \renewcommand{\arraystretch}{1.3}
    \begin{tabular}{l l}
        \toprule
        \rowcolor{gray!15}
        \textbf{Input} & \textbf{\texttt{ldr r1, r2}} \\
        \midrule
        \rowcolor{gray!15}
        \textbf{Tokenizer} & \textbf{Tokens} \\
        \midrule
        DeepSeek/Yi-Coder & 
        \begin{tabular}[t]{@{}l@{}}
            \colorbox{red!15}{\strut\texttt{ld}}%
            \colorbox{yellow!15}{\strut\texttt{r}}%
            \colorbox{green!15}{\strut\texttt{\char32}}%
            \colorbox{blue!15}{\strut\texttt{r}}%
            \colorbox{yellow!15}{\strut\texttt{1}}%
            \colorbox{orange!15}{\strut\texttt{,}}%
            \colorbox{purple!15}{\strut\texttt{\char32}}%
            \colorbox{cyan!15}{\strut\texttt{r}}%
            \colorbox{yellow!15}{\strut\texttt{2}}
        \end{tabular} \\[2ex]
        Our Extended Tokenizer & 
        \begin{tabular}[t]{@{}l@{}}
            \colorbox{red!15}{\strut\texttt{ldr}}%
            \colorbox{green!15}{\strut\texttt{\char32}}%
            \colorbox{blue!15}{\strut\texttt{r1}}%
            \colorbox{orange!15}{\strut\texttt{,}}%
            \colorbox{purple!15}{\strut\texttt{\char32}}%
            \colorbox{cyan!15}{\strut\texttt{r2}}
        \end{tabular} \\
        \bottomrule
    \end{tabular}
    \caption{Comparison of tokenization approaches between DeepSeek/Yi-Coder and our extended tokenizer. Spaces are represented as \texttt{\char32} and shown with colored backgrounds to highlight token boundaries. Note how our tokenizer groups related tokens (e.g., \texttt{ldr} and \texttt{r1}) as singular units.}
    \label{tab:tokenizer_comparison}
\end{table}
\vspace{-0.1in}

To enhance our LLMs' understanding and generation of assembly code, we extended the tokenizer to include the most frequently used opcodes and register names from both x86 and ARM architectures (see Table~\ref{tab:tokenizer_comparison}). Tokenization segments raw text into tokens for the model to process~\cite{transformers}, allowing accurate recognition of assembly language components. This customization efficiently represents the distinct semantics of each instruction set, improving the model's ability to parse and generate correct translations by aligning with the low-level details of the input assembly code.

\section{Experiments and Evaluation} \label{eval}

We present a comprehensive evaluation of CRT's effectiveness in x86-to-ARM binary transpilation across three dimensions: training data preparation (\S\ref{subsec:train-data}), hyperparameter tuning (\S\ref{subsec:setup}), and model architecture selection. Using standard benchmarks (\S\ref{sec:eval-bench}) and metrics (\S\ref{sec:metrics}), we assess the generated assembly code's semantic preservation and functional correctness.

\subsection{Training Data}\label{subsec:train-data}

The training dataset was derived from AnghaBench~\cite{anghabench}, a comprehensive benchmark suite containing 1 million compilable C programs mined from major public C repositories on GitHub. From this benchmark, we randomly sampled 500k programs to form our training set, equivalent to 8 billion tokens. These programs were then compiled to x86 using \texttt{gcc}~\cite{gcc} and cross-compiled to ARMv5 using \texttt{ARM-gnueabi-gcc}~\cite{gnueabi}; both sets were generated on an AMD Ryzen 7 processor. 

\subsection{Experimental Setup} \label{subsec:setup}
All of our hyperparameter optimization experiments were conducted on a small 100k portion of AnghaBench. We tested various hyperparameter settings on this subset of our benchmark. After identifying the optimal configuration, we scaled up the training data to 500k samples. We trained three models: DeepSeek-Coder1.3B~\cite{deepseek-coder}, Yi-Coder2B~\cite{yicoder}, and BART-Large (300M)~\cite{bart} on the AnghaBench dataset. Given the dataset size of 1 million samples, with an average of 13k tokens per sample, we opted for smaller models and worked with reduced dataset size of 500k samples. All models were trained using four A100 GPUs (40 GB each). Training with 500k samples, a batch size of 4, and 2 epochs required three days. To conserve memory, mixed precision training with \texttt{bfloat16} was employed. Given limited capacity for large batch sizes, we applied gradient accumulation~\cite{gradient-accumulation} with an effective batch size of 4. Additionally, all models were trained with optimization level \verb|-O0|, and we used paged AdamW~\cite{adamw} to avoid memory spikes, with a weight decay of 0.001. We chose a small learning rate of $1 \times 10^{-4}$ with a linear schedule, as experiments indicated this schedule performed best. All models were trained with a context window of 16k. 

For inference, we used caching to enhance inference speed and disabled sampling to ensure deterministic outputs. Our evaluation set was sourced from the LLM4Decompile~\cite{tan2024llm4decompile} set, compiled to x86. We employed QEMU~\cite{qemu} to simulate the evaluation environment. Following training, we apply quantization techniques (e.g., \texttt{bfloat16}, \texttt{int8}, and \texttt{int4}) using \texttt{llama.cpp}~\cite{llama-cpp} to optimize for efficient inference on CPU-based devices. This step is crucial to maintain high performance while reducing the computational load, making the solution feasible for local deployment. 

\subsection{Evaluation Benchmark} \label{sec:eval-bench}

CRT's performance accuracy is evaluated using the HumanEval benchmark, originally introduced by \cite{chen2021evaluating} for Python code generation. The benchmark consists of 164 programming problems that assess language comprehension, reasoning, and algorithmic thinking. For our evaluation, we utilize the C-translated version from LLM4Decompile \cite{tan2024llm4decompile}, which maintains the same problems while converting both function implementations and test cases to C code.

To ensure thorough testing, we measure code line coverage using \texttt{gcov}, GNU's source code coverage analysis tool. As emphasized by \cite{myers2011art}, line coverage is a fundamental metric in software testing that indicates which lines of code were executed at least once during testing, helping identify untested code paths and potential blind spots in test suites. The higher the line coverage percentage, the more comprehensive the testing. HumanEval resulted in an average line coverage of 98.81\%, indicating that nearly all lines of code were executed during testing.

For the evaluation process, we generate the corresponding assembly code pairs following the training data preparation process detailed in Section \S\ref{subsec:train-data}.

\subsection{Evaluation Metrics} \label{sec:metrics}
We evaluate our approach using two primary metrics:

\textbf{Edit Distance: }Following prior work~\cite{lee2024guess}, we employ the Levenshtein edit distance \cite{lcvenshtcin1966binary} between the ground truth and transpiled ARM assembly as an initial measure of syntactic similarity. However, this metric is limited because semantically equivalent assembly sequences can differ syntactically due to variations like register allocation or instruction ordering (\S\ref{discussion}).  

\textbf{Functional Correctness:} 
To overcome the limitations of edit distance, we assess the functional equivalence by running the ground truth and transpiled code against comprehensive test suites, employing software testing principles via unit test coverage. A transpilation is considered correct if it passes all test cases of the corresponding program. The experimental results presented in~\S\ref{sec_results} show that a significant portion of the transpilation results differ syntactically (having non-zero edit distance) from the ground truth yet preserve program semantics and execute correctly.

\begin{table*}[t]
    \centering
    \small
    \renewcommand{\arraystretch}{1.2}
    \begin{tabular}{l c c c}
        \toprule
        \textbf{Model} & \textbf{Average Edit Distance} ($\downarrow$) & \textbf{Exact Match} ($\uparrow$) & \textbf{Test Accuracy} ($\uparrow$) \\
        \midrule
        GPT4o~\cite{gpt4o} & 1296 & 0\% & 8.18\% \\
        DeepSeekCoder2-16B~\cite{deepseekcoder2} & 1633 & 0\% & 7.36\% \\
        Yi-Coder-9B~\cite{yicoder} & 1653 & 0\% & 6.33\% \\
        \midrule
        Yi-coder-1.5B & 275 & 16.98\% & 49.69\% \\
        DeepSeekCoder-1.3B & 107 & 45.91\% & 77.23\% \\
        DeepSeekCoder-1.3B-xTokenizer-int4 & 119 & 46.54\% & 72.96\% \\
        DeepSeekCoder-1.3B-xTokenizer-int8 & \textbf{96} & 49.69\% & 75.47\% \\
        DeepSeekCoder-1.3B-xTokenizer & 165 & \textbf{50.32\%} & \textbf{79.25\%} \\
        \bottomrule
    \end{tabular}
    \caption{Comparison of models' performance on the \textit{x86 to ARM} transpilation task, measured by Edit Distance, Exact Match, and Test Accuracy. The top portion lists pre-existing models, while the bottom portion lists models trained by us. Arrows~($\uparrow$,~$\downarrow$) indicate whether higher or lower values are better for each metric. The best results are highlighted in \textbf{bold}.}
    \label{tab:results}
\end{table*}

\section{Results} \label{sec_results}

We evaluate the efficacy of our transpiler for CISC-to-RISC assembly translation, focusing on the correctness of the output ARM assembly. Utilizing the metrics defined above (\S\ref{eval}), we compare our approach with state-of-the-art coding LLMs and evaluate our approach for x86 to ARM transpilation (Table\ref{tab:results}). 

\subsection{Transpiler Validation}
Of the evaluated LLMs, DeepSeekCoder-1.3B-xTokenizer (our model utilizing the tokenizer and trained as described in \S\ref{subsec:setup}) achieves a test accuracy of \textbf{79.25\%} on x86 to ARM transpilation, substantially outperforming larger models such as GPT4o (8.18\%), DeepSeekCoder2-16B (7.36\%), and Yi-Coder-9B (6.33\%) (Table~\ref{tab:results}). Despite being 9$\times$ to 20$\times$ smaller in size, our model exhibits at up to 9.8$\times$ the accuracy, highlighting the effectiveness of our approach.

Compared to prior work on ARM to RISC-V translation ~\cite{lee2024guess} (which achieves 68\% accuracy - 80\% when using symbolic analysis - with a similarly sized model), our model attains higher accuracy on the more complex task of x86 to ARM translation without the use of external, symbolic testing or tools. We attribute this approximate 7.8\% difference to our better tokenization scheme (which is needed for the CISC/RISC disambiguation) as well as the 8x larger context window. 

An interesting comparison arises between DeepSeekCoder-1.3B and Yi-Coder-1.5B; DeepSeekCoder outperforms Yi-Coder by 28\%. We attribute this to DeepSeekCoder being trained from scratch on 2 trillion code tokens, whereas Yi-Coder is based on the Yi-chat model, potentially limiting its efficacy for this specific task. 

Additionally, the results from our aligned BART shows a 12\% jump in accuracy from baseline BART inspired by~\cite{lee2024guess}. This shows that our alignment technique is effective even with a very low context window of 1024 but yet still lacks behind large context window of DeepSeekCoder by 62.8\%. 

Our model exhibits strong syntactic robustness, as evidenced by the high number of transpiled programs with an edit distance of zero (Table~\ref{tab:results}, column~3). This indicates that the generated assembly code is often syntactically identical to the reference code, demonstrating proficiency in producing correct assembly language syntax for both RISC architectures (ARM \& RISC-V). Moreover, the absence of syntax errors allows us to focus our evaluation on functional correctness, measured by test accuracy metrics. This aligns with our observation that LMs rarely produce typos or grammatical errors, even in assembly instructions, based on our tests.

\subsection{Impact of Tokenizer and Quantization}

Enhancements in our tokenizer contribute to the model's accuracy. DeepSeekCoder-1.3B-xTokenizer increases accuracy by 2\% compared to DeepSeekCoder-1.3B, indicating that our extended tokenizer enables more efficient learning. Additionally, the optimized tokenizer reduces the average number of tokens by 7.27\%, improving inference speed.

Our models quantized to \texttt{int8} and \texttt{int4} precision achieve comparable results to the \texttt{float32} models. For ARM transpilation, accuracy decreases by only 3.8\% when moving from \texttt{float32} to \texttt{int8}, and by nearly 2.5\% when moving to \texttt{int4} (Table~\ref{tab:results} column~2).   

\subsection{Training Performance}

Selecting appropriate training hyperparameters has a significant impact on model accuracy and edit distance. In our experiments, implementing warmup steps~\cite{warmup} notably improved the model's edit distance, decreasing it by 8\%. This improvement is attributed to enhanced convergence and training stability, which are crucial for complex tasks like x86 to ARM transpilation. Since we are fine-tuning models that were initially trained to follow specific instructions to produce code, the use of warmup steps with a higher initial learning rate helps our model transition from this instruction-based generation setting to our transpilation setting, thereby avoiding early overfitting and increasing generalization~\cite{why-warmup}.

\subsection{Inference Performance}

\begin{figure*}[t]
   \centering
   \small
   \begin{subfigure}[b]{0.30\textwidth}
       \includegraphics[width=\linewidth]{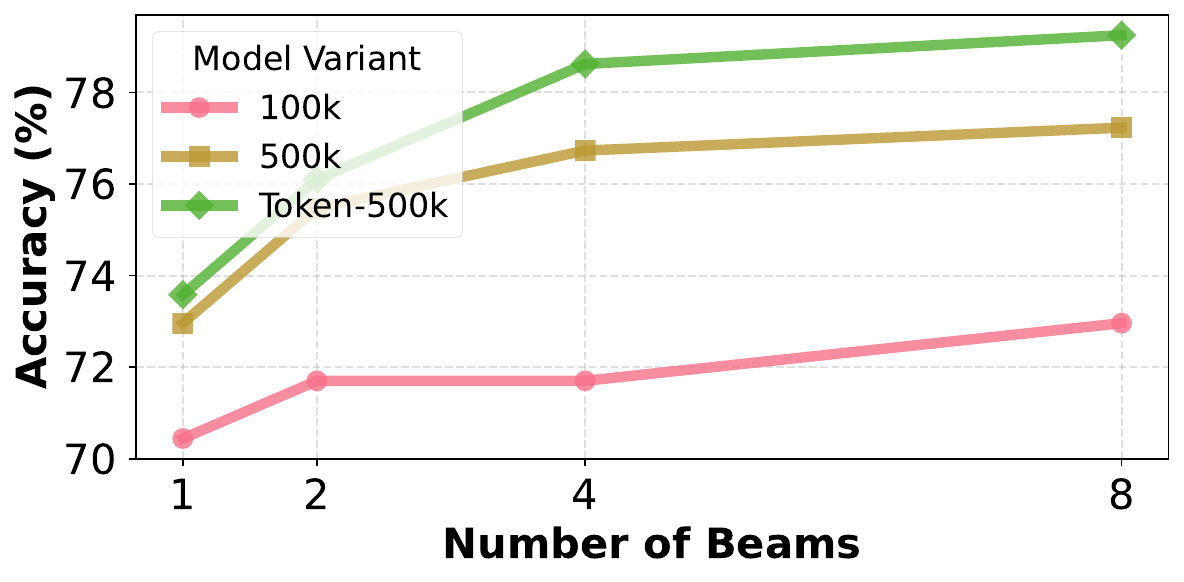}
       \caption{}  
       \label{fig:beams-vs-accuracy}
   \end{subfigure}
   \hfill
   \begin{subfigure}[b]{0.30\textwidth}
       \includegraphics[width=\linewidth]{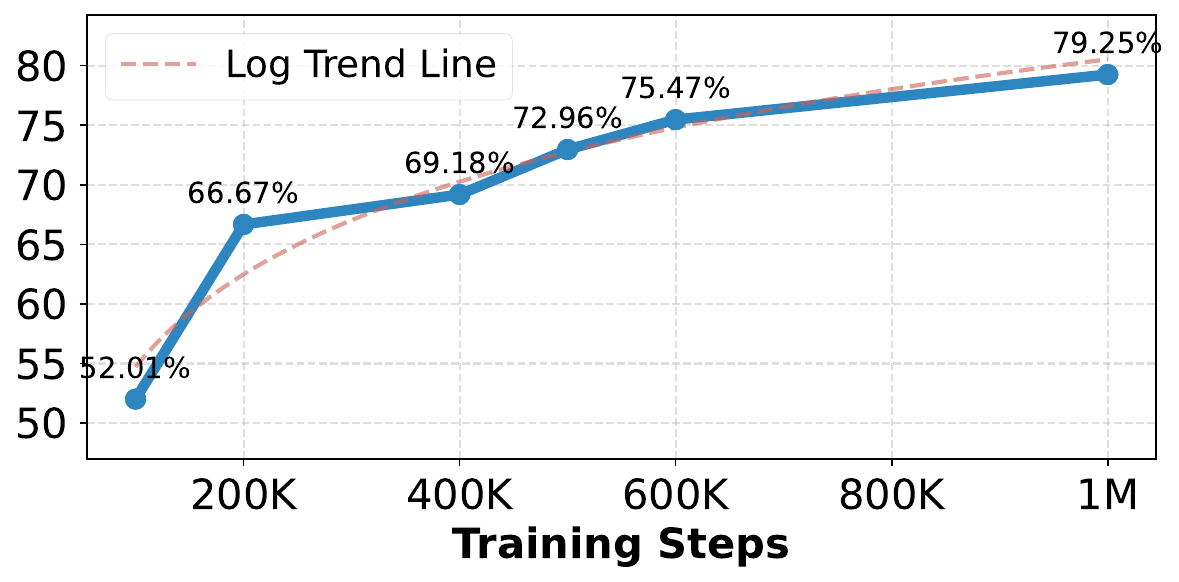}
       \caption{}  
       \label{fig:data-size-vs-accuracy} 
   \end{subfigure}
   \hfill
   \begin{subfigure}[b]{0.30\textwidth}
       \includegraphics[width=\linewidth]{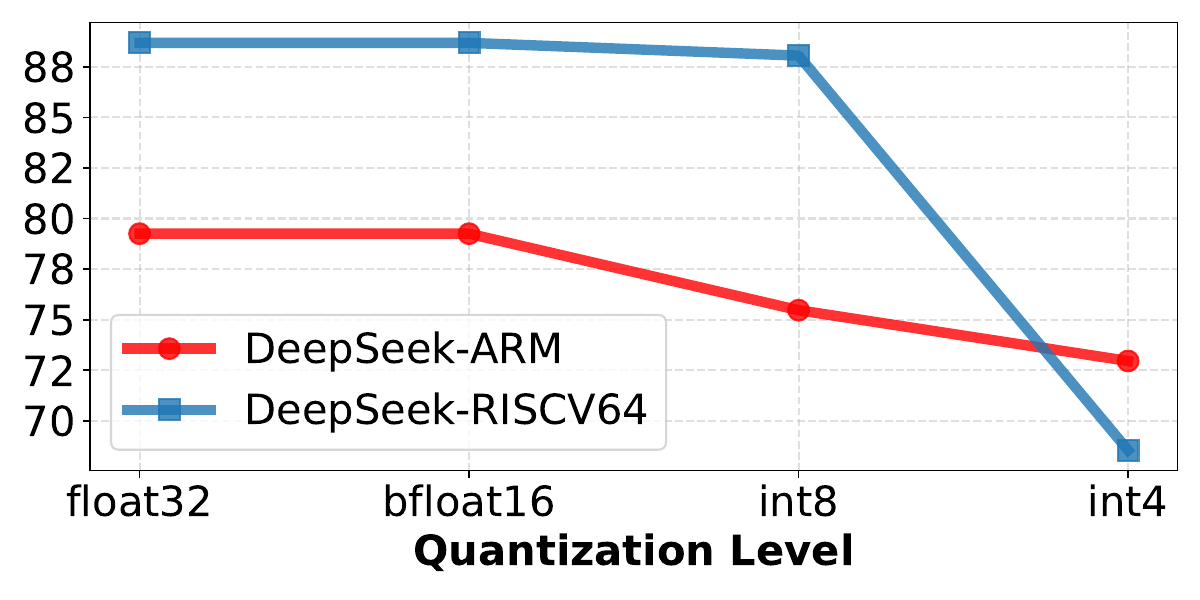}
       \caption{}  
       \label{fig:quantization-vs-accuracy} 
   \end{subfigure}
   \vspace{-1em}
   \caption{DeepSeek-1.3B performance: (a) Accuracy across beam sizes (1, 2, 4, 8) for different training data sizes. (b) Accuracy progression over training steps with a logarithmic trend. (c) Quantization impact (float32, bfloat16, int8, int4) on ARM and RISC-V64.}
   \label{fig:deepseek-analysis}
\end{figure*}

As shown in Figure~\ref{fig:deepseek-analysis}, we analyze the model's accuracy across different inference configurations. Our beam search experiments (Figure~\ref{fig:beams-vs-accuracy}) demonstrate that increasing the number of beams improves accuracy. Specifically, as the beam size increases from 1 to 8, the model explores multiple decoding paths, akin to human-like consideration of different options. This approach is similar to the symbolic solving post-processing technique used in~\cite{lee2024guess}, but with a computational complexity of $O(bN^2)$ compared to their $O(2^N)$ complexity. This increased flexibility allows the model to produce more robust solutions. However, larger beam sizes increase computational overhead, presenting a trade-off between performance and inference speed.

From a runtime performance, our models exhibit real-world inference efficiency. On an NVIDIA A100 GPU with 15.6~TFLOPS, our model generates a sample with a 16k context length in 18.3 seconds on average, achieving a rate of 437.2 tokens per second. On a Ryzen 7 CPU, the model achieves 18.51 tokens per second with \texttt{int8} quantization and 87.23 tokens per second with \texttt{int4} quantization using Ollama, demonstrating applicability for real-world use cases requiring efficient inference.

Figure~\ref{fig:data-size-vs-accuracy} shows that our model's accuracy increases logarithmically with the size of the training dataset. As more data is provided, the model better learns to reason about transpilation with diverse training examples following LLMs scaling laws~\cite{scaling-laws}.

We also experimented with different compilers (GCC and Clang) during training and found that both produced similar results, with an edit distance of 133 and exact match accuracy of 50.31\%. All results are thus reported with the GCC compiler, due to its compatibility with QEMU for running tests as mentioned in section~\ref{subsec:setup}. 

\begin{table*}[t]
    \centering
    \small
    \renewcommand{\arraystretch}{1.2}
    \begin{tabular}{l c c c}
        \toprule
        \textbf{Model} & \textbf{Average Edit Distance} ($\downarrow$) & \textbf{Exact Match} ($\uparrow$) & \textbf{Test Accuracy} ($\uparrow$) \\
        \midrule
        GPT4o~\cite{gpt4o} & 1293 & 0\% & 7.55\% \\
        DeepSeekCoder2-16B~\cite{deepseekcoder2} & 1483 & 0\% & 6.29\% \\
        \midrule
        DeepSeekCoder-1.3B-xTokenizer-int4 & 112 & 14.47\% & 68.55\% \\
        DeepSeekCoder-1.3B-xTokenizer-int8 & 31 & 69.81\% & 88.05\% \\
        DeepSeekCoder-1.3B-xTokenizer & \textbf{27} & \textbf{69.81\%} & \textbf{88.68\%} \\
        \bottomrule
    \end{tabular}
    \caption{Comparison of models' performance on the \textit{x86 to RISCv64} transpilation task.}
    \label{tab:results-risc}
\end{table*}

\subsection{Analysis of Transpiled Assembly Code}
\label{sub:analysis}

To better understand the transpilations generated by CRT, we examined the evaluation benchmark's ground truth and compared it with the transpiled code. An interesting case arises when the transpiled code is correct, but the edit distance from the ground truth deviates from zero.

Our investigation revealed various patterns of implementation that maintain functional correctness despite differing from the ground truth. Commutative operations, like the \code{add} opcode, sometimes show order differences between the ground truth and predicted code. Register allocation variations occur when different registers are chosen for the same operations, as long as data flow and register dependencies are preserved. Memory location swapping is another pattern where variables are stored in different memory locations, yet their relationships remain consistent. Additionally, stack frame sizes sometimes differ between the ground truth and transpiled code.

We also identified cases of instruction-level semantic equivalence, where different instruction sequences achieve the same logical result. For example, multiplication in the ground truth might use a direct \code{mul} opcode, while the transpiled code achieves the same computation using shifts and additions. Variations in constant handling were also noted, with the transpiled code sometimes using immediate values directly in instructions while the ground truth loads constants from memory. Instruction consolidation is common, with the transpiled code combining multiple instructions (\code{mov} \reg{r1}, \reg{r2}; \code{add} \reg{r1}, \reg{r1}, \imm{\#1}) into single, streamlined versions (\code{add} \reg{r2}, \reg{r2}, \imm{\#1}) that maintains functionality. In many instances, these patterns appear together in what we refer to as \text{"}composite variations\text{"}.

The error patterns become particularly evident in assembly code, where even slight variations can lead to significant functional issues. For instance, non-commutative operations are prone to critical errors when operand order is reversed, as seen in subtraction, where using \code{sub} \reg{r1}, \reg{r3}, \reg{r2} instead of \code{sub} \reg{r1}, \reg{r2}, \reg{r3} completely changes the computation. Incorrect register management, where registers are overwritten prematurely, can lead to data loss. Immediate value errors also appear, particularly in shift operations; for instance, \code{asr} \reg{r2}, \reg{r2}, \imm{\#1} is mistakenly used for division by 2 when division by 4 (\code{asr} \reg{r2}, \reg{r2}, \imm{\#2}) is needed, causing computational discrepancies. Memory addressing errors, such as misaligned access, occur when values are stored and retrieved from incorrect offsets, resulting in data corruption. Additional examples are provided in the Appendix\S\ref{subsec:assembly_analysis}.

We also observe for incorrect cases with a high edit distance, the transpiled code diverges significantly from the ground truth, often reflecting substantial deviations from the intended logic---sometimes even resulting in unintended behaviors like infinite loops. Such drastic variations make it difficult to interpret or troubleshoot the transpiled output by simply comparing it to the ground truth, as the underlying logic no longer aligns.
\subsection{Extension to RISC-V ISA}

To demonstrate the generality of our method, we also trained our model on the task of transpiling from x86 to RISC-V64, achieving a test accuracy of \textbf{88.68\%} (Table~\ref{tab:results-risc}). Notably, our model significantly outperforms existing models like GPT4o and DeepSeekCoder2-16B, which achieved much lower test accuracies of 7.55\% and 6.29\%, respectively. However, this result indicates that transpiling to RISC-V64 is easier than to ARM, possibly due to RISC-V's simpler and more consistent instruction set compared to the more complex instructions in recent ARM architectures. For RISC-V efficient transpilation (Table~\ref{tab:results-risc} column~2), the \texttt{int8} quantized model's accuracy is nearly identical to the \texttt{float32} model, with a minor drop of 0.63\% and the same exact match accuracy. However, the \texttt{int4} quantized model experiences a significant accuracy drop of 19.5\%. The results not only reinforce that we can deploy our model on consumer-grade machines with high-efficiency, but also from an information theory perspective that the number bits necessary for CISC-to-RISC transpilation can still accurately encode functional semantics for machine code. 

\section{Discussion} \label{discussion}

Our proposed model effectively tackles the challenges of CISC-to-RISC assembly code transpilation, achieving high performance without extensive model scaling. In this section, we provide insights into our model's behavior and the factors influencing its performance.

One key observation is the performance difference between x86-to-ARM and x86-to-RISC-V64 transpilation. Our model performs better when translating to RISC-V, likely due to its simpler instruction set (47 instructions) compared to ARM's more complex set (approximately 100 instructions), as shown in Figure~\ref{fig:assembly-comparison}. This suggests that the complexity of the target ISA significantly affects the difficulty of the transpilation task. Consequently, we focused most of our experiments on x86-to-ARM transpilation to address challenges associated with more complex RISC ISAs.

Another important aspect is the model's syntactic flexibility. While CRT achieves high functional correctness, it often produces syntactically diverse outputs, as indicated by the high edit distance metric. Instead of memorizing patterns, the model reasons through register mappings and instruction sequences, generating functionally correct but syntactically varied code. This flexibility is beneficial in real-world code generation, where functional equivalence is more valuable than syntactic similarity. For example, the model may choose different sets of registers for each instance, as long as it maintains consistency in their usage. Acceptable variations include using different operand orders in commutative instructions (e.g., \code{add} \reg{r0}, \reg{r1}, \reg{r2} vs.\ \code{add} \reg{r0}, \reg{r2}, \reg{r1}, assigning temporary variables to different registers, combining multiple instructions into a more efficient single instruction, varying allocated memory size or resource usage without affecting functionality, employing different methods for loading constants, and consistently swapping memory locations for variables (e.g., using \texttt{[\code{fp}, \imm{\#-8}]} instead of \texttt{[\code{fp}, \imm{\#-12}]} for a loop counter). These variations showcase the model's ability to generate functionally equivalent but syntactically diverse code, which is often desirable in assembly code generation.

Despite these successes, CRT is not without limitations. We observed that certain errors persist, which can be categorized into three main types (1) \textbf{Register Allocation Errors (14.11\%):} Allocating registers that have already been allocated, leading to memory issues (2) \textbf{Addressing Errors (62.03\%):} Jumping to prohibited memory addresses or copying incorrect addresses from x86 code (3) \textbf{Other Errors (23.86\%):} Including invalid constants and floating-point exceptions. We noticed that additional training reduced the number of Type 1 errors by 21.45\%, Type 2 errors by 7.67\%, and Type 3 errors by 1.39\%. This suggests that while increased training data can mitigate some errors, particularly those related to register allocation, other errors may require different strategies. 

One of the challenges we identified is related to the model's handling of numerical tokens. Current tokenizers often treat each digit of a number as a separate token, which can hinder the model's ability to correctly process long numerical values, such as memory addresses or constants that need to be accurately copied or slightly modified from the x86 code. We believe that developing a better tokenizer that handles numbers as single tokens could significantly improve the model's performance, particularly in reducing addressing and constant-related errors. 

Notably, our model, despite being at least seven times smaller than those used in similar studies~\cite{lee2024guess,tan2024llm4decompile}, was effective in this task. This suggests that for machine language translation, model quality and data curation are more critical than sheer model size. As shown in Figure~\ref{fig:data-size-vs-accuracy}, the quality and size of the training data significantly impact performance. Additionally, using an extended tokenizer improved the understanding of instructions and registers, and a longer context window enabled the model to track register usage effectively. Achieving these results with a small, quantized model indicates that efficiency and practicality need not be sacrificed for performance. This has significant implications for deploying such models on resource-constrained hardware common in embedded systems and other RISC applications.

Looking ahead, there are several avenues for improving our model. In addition to developing a better tokenizer, as previously mentioned, incorporating more diverse and comprehensive training data could further reduce errors, particularly those related to addressing and constants. Exploring techniques such as incorporating domain-specific knowledge or constraints into the model training process might also enhance performance.

Overall, our findings highlight that high-quality assembly code transpilation across diverse ISAs requires a holistic approach that goes beyond merely increasing model size. Thoughtful design of tokenization and training processes, attention to ISA-specific challenges, and efficient quantization collectively enable high-performing, deployment-ready models. These insights contribute to a deeper understanding of how models can be designed to facilitate the industry's ongoing transition towards scalable, energy-efficient processor architectures, supporting software compatibility and performance across platforms.

\begin{table}[b]
  \centering
  \small
  \renewcommand{\arraystretch}{1.2}
  \begin{tabular}{l r r r}
      \toprule
      \small
      \textbf{Model} & \textbf{AED ($\downarrow$)} & \textbf{EM ($\uparrow$)} & \textbf{Acc. ($\uparrow$)} \\
      \midrule
      CRT ARMv5 & 165 & 50.32\% & 79.25\% \\
      \midrule
      CRT ARMv8 & 105 & 50.61\% & 75.0\% \\
      \bottomrule
  \end{tabular}
  \caption{Correctness comparison between CRT implementations on ARMv5 and ARMv8 architectures. Metrics include Average Edit Distance (AED), Exact Match (EM), and Test Accuracy (Acc.).}
  \label{results-armv8}
\end{table}

\section{Case Study} 
\label{study}

\begin{figure*}[t]
    \centering
    \begin{subfigure}[b]{0.32\textwidth}
        \includegraphics[width=\linewidth]{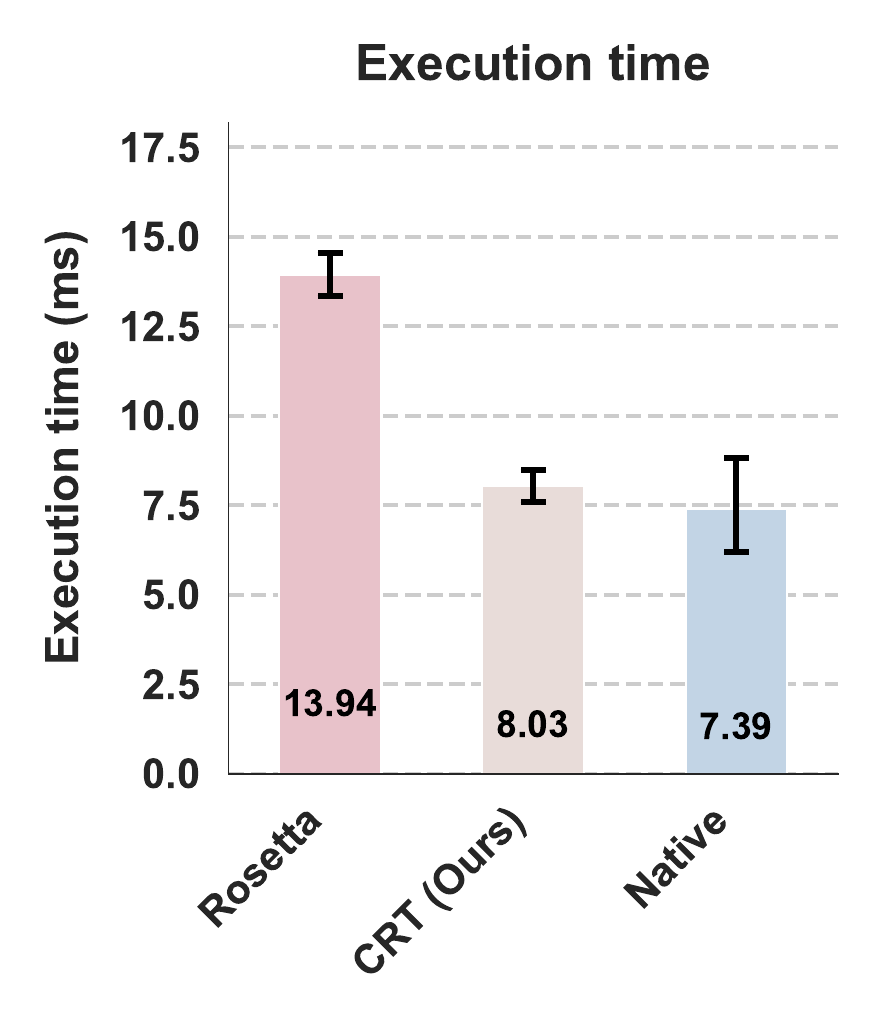}
        \label{fig:execution_time}
        
    \end{subfigure}
    \hfill
    \begin{subfigure}[b]{0.32\textwidth}
        \includegraphics[width=\linewidth]{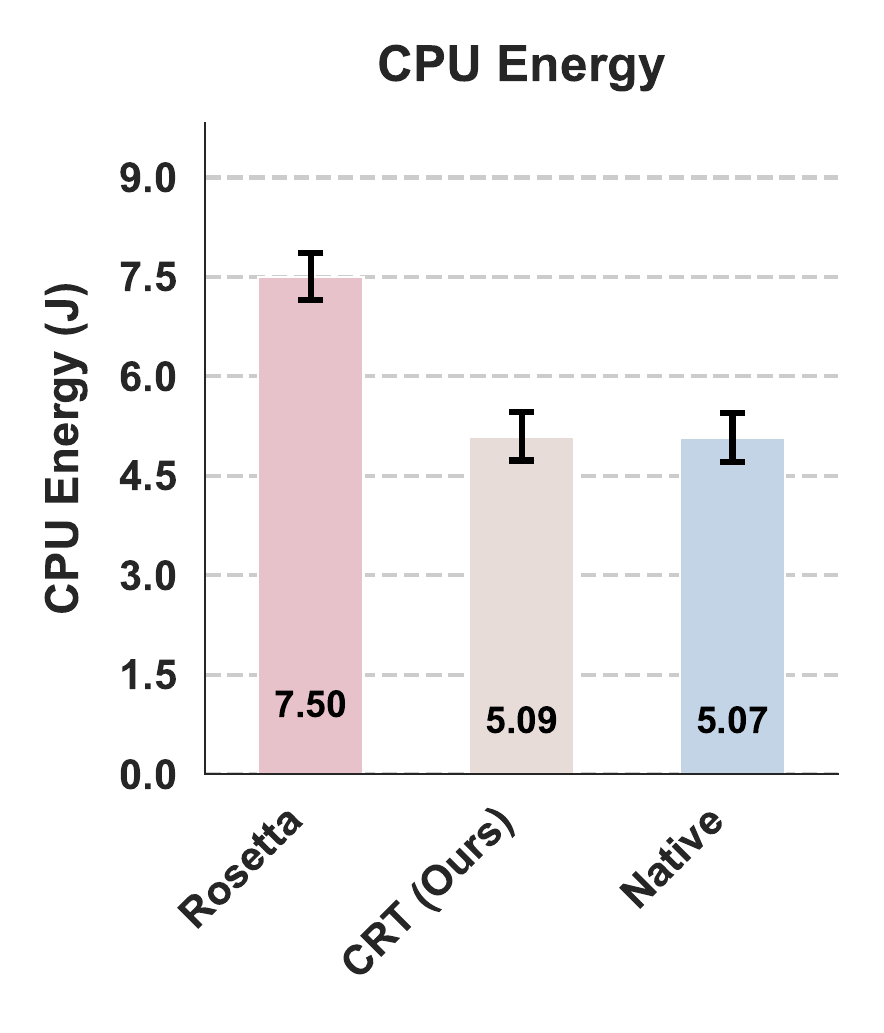}
        \label{fig:cpu_energy}
        
    \end{subfigure}
    \hfill
    \begin{subfigure}[b]{0.32\textwidth}
        \includegraphics[width=\linewidth]{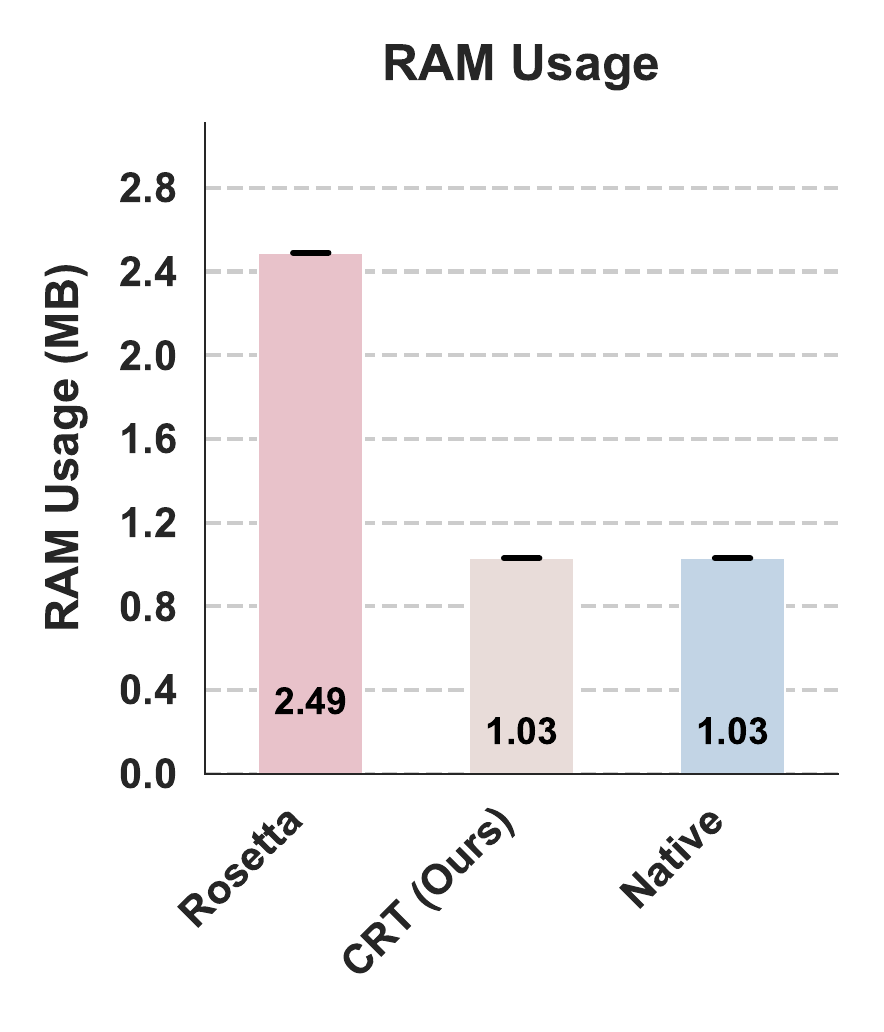}
        \label{fig:ram_usage}
        
    \end{subfigure}
    \vspace{-0.4in} 
    \caption{Measured execution time, energy utilization, and RAM usage across different settings on Apple M2 Macbook.}
    \label{fig:performance_comparison}
\end{figure*}

To evaluate the efficiency of our transpiler, we performed a real world study on an Apple M2 Pro machine, which features a more recent version of ARM, the ARM64v8-A architecture. This transition offered two key advantages: first, it enabled us to use the native ARM compiler toolchain to generate the target instruction sequences, rather than relying on cross-compilation; second, through Apple's Rosetta 2 translation layer, we established a unified testing environment enabling comparative analysis across different execution modes within the same hardware platform.

\subsection{Experimental Setup}
\label{case_study_setup}
The evaluations were conducted using Apple Clang 14.0.3 (clang-1403.0.22.14.1) on an Apple M2 Pro processor with 16GB of RAM running macOS 13.7. All programs were compiled targeting the arm64-apple-darwin22.6.0 architecture with \verb|-O0| optimization level. Additional details are the same as \S\ref{subsec:setup} for our transpilation correction evaluation.

We examine the execution characteristics across three distinct execution environments on the M2 Pro machine. First, we establish the baseline performance by executing natively compiled ARM64 binaries. Second, we measure the performance of x86 binaries executed through Apple's Rosetta 2 dynamic binary translation layer. Third, we evaluate our CRT transpiled code that directly transforms x86 assembly to ARM64. For each environment, we analyze three performance dimensions: execution time, CPU energy consumption measured using powermetrics instrumentation, and memory utilization patterns. To ensure statistical validity, we execute each program 100 times and compute its geometric mean for reporting~\cite{fleming1986not}. All performance evaluations are conducted under controlled conditions.

The functional correctness of the proposed approach, when trained on ARMv8, yielded the results shown in Table \ref{results-armv8}. A key finding is that compared to ARMv5, we observed a performance drop of 4.25\% with ARMv8 achieving 75.0\% on the evaluation-set. This decline can be attributed to ARMv8's increased architectural complexity (refer to Appendix~\ref{app-compare-archs}). The increased complexity is evident in several areas, including register usage patterns, addressing modes, comparison operations, and floating-point handling. In each of these areas, ARMv8 adopts more sophisticated approaches compared to ARMv5's straightforward implementations. These architectural features make instruction patterns more challenging for the LLM to learn effectively.

\vspace{-.1in}
\begin{figure}[!h]
    \centering
    \small
        \includegraphics[width=0.9\columnwidth]{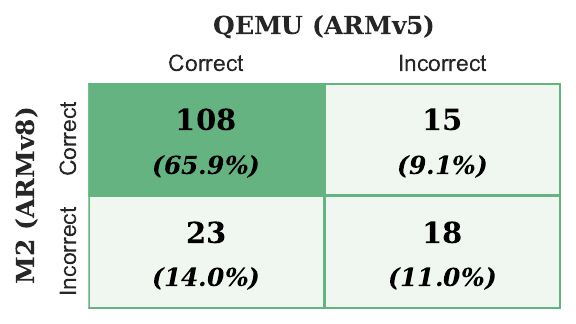}
    \vspace{-.1in}
    \caption{Confusion matrix of the proposed approach executed on QEMU (ARMv5) and M2 (ARMv8) for HumanEval programs.}
    \label{confusion-matrix}
\end{figure}

Examining the confusion matrix (Figure~\ref{confusion-matrix}) for the proposed approach's performance across ARMv5 and ARMv8 architectures reveals an agreement of 76.8\%. However, we observe distinct failure patterns (15 versus 23 cases failing uniquely on ARMv5 and ARMv8 respectively), suggesting architecture-specific performance variations that induce divergent error modes in the LLM's execution.

\subsection{Performance Analysis}
The performance evaluation shows that CRT achieves near-native efficiency across multiple metrics. With execution times nearly matching native code, CRT delivers a 1.73x speedup over Rosetta, along with 1.47x better energy efficiency and 2.41x better memory efficiency. Notably, CRT’s memory footprint remains close to native execution (1.034 MB vs. 1.03 MB), whereas Rosetta requires 2.49 MB.

\subsection{ARMv5 versus ARMv8 Analysis}

The proposed transpiler enables efficient cross-architecture code execution with performance nearing native compilation, proving the feasibility of LLM-based binary translation. These results indicate that machine learning approaches can effectively learn complex mappings between instruction set architectures, maintaining high performance with significant improvements over Rosetta.

Looking forward, this opens the door for efficient CISC to RISC transitions, particularly to ARM architectures, while maintaining seamless control over legacy x86 software with reduced overhead and improved performance. This could potentially accelerate the adoption of ARM architectures across enterprise environments.

\vspace{-0.05in}
\section{Conclusion}
This paper presents CRT, a language model-guided transpiler for converting CISC (x86) assembly code to RISC (ARM and RISC-V) architectures. Our work explores a shift in systems software design - moving away from hand-crafted rules and predetermined translation patterns toward more intelligent and adaptable solutions. As hardware architectures continue to diversify and evolve, particularly with the rise of domain-specific accelerators, learning-based approaches may become increasingly relevant for maintaining software portability while maximizing hardware performance. The future of systems software lies in leveraging machine learning to create solutions that can adapt to architectural changes without requiring separate engineering efforts for each new target platform.

\nocite{langley00}
\clearpage
\bibliography{references}
\bibliographystyle{mlsys2024}
\clearpage
\appendix

\appendix

\section{Appendix} \label{appendix}

\subsection{Analysis of Assembly Code Variations}
\label{subsec:assembly_analysis}

In this section, we present a detailed analysis of the variations between ground truth and predicted assembly code from our evaluation benchmark. Our analysis focuses on two key aspects: (1) functionally equivalent code with non-zero edit distance, and (2) incorrect implementations that fail the test cases. For functionally equivalent code, we identified several common patterns that maintain correctness despite syntactic differences. These patterns range from simple variations(Table~\ref{tab:simple_variations}) like register allocation and commutative operations, to complex patterns that combine multiple types of differences (Table~\ref{tab:complex_variations}).

\begin{table}[!h]
\large
\centering
\caption{Simple Variation Patterns in Functionally Equivalent Code}
\label{tab:simple_variations}
\resizebox{\linewidth}{!}{\begin{tabular}{@{}p{0.8cm}p{0.8cm}p{9.9cm}@{}}
\toprule
\textbf{Prog ID} & \textbf{Edit Dist} & \textbf{Example} \\
\midrule
P75 & 8 & \textbf{Operands in arithmetic operations can be reordered if operation is commutative} \\
& & Ground truth: \code{add} \reg{r1}, \reg{r2}, \reg{r3} \\
& & Predicted: \code{add} \reg{r1}, \reg{r3}, \reg{r2} \\
\midrule
P108 & 16 & \textbf{Different registers can be chosen for temporary values while maintaining same data flow} \\
& & Ground truth: \code{mov} \reg{r2}, \reg{r0}; \code{add} \reg{r2}, \reg{r2}, \imm{\#1} \\
& & Predicted: \code{mov} \reg{r3}, \reg{r0}; \code{add} \reg{r3}, \reg{r3}, \imm{\#1} \\
\midrule
P8 & 12 & \textbf{Local variables can be stored at different stack locations while maintaining correct access patterns} \\
& & Ground truth: \code{str} \reg{r1}, \mem{[fp, \#-8]}; \code{str} \reg{r2}, \mem{[fp, \#-12]} \\
& & Predicted: \code{str} \reg{r1}, \mem{[fp, \#-12]}; \code{str} \reg{r2}, \mem{[fp, \#-8]} \\
\midrule
P119 & 6 & \textbf{Compiler-generated symbol names can differ while referring to same data} \\
& & Ground truth: \code{.word} \code{out.4781} \\
& & Predicted: \code{.word} \code{out.4280} \\
\midrule
P135 & 12 & \textbf{Multiple instructions can be combined into single equivalent instruction} \\
& & Ground truth: \code{mov} \reg{r3}, \reg{r0}; \\
& & \hspace{2.50cm}\code{str} \reg{r3}, \mem{[fp, \#-8]} \\
& & Predicted: \code{str} \reg{r0}, \mem{[fp, \#-8]} \\
\midrule
P162 & 4 & \textbf{Stack frame offsets can vary while maintaining correct variable access} \\
& & Ground truth: \code{strb} \reg{r3}, \mem{[fp, \#-21]} \\
& & Predicted: \code{strb} \reg{r3}, \mem{[fp, \#-17]} \\
\midrule
P88 & 23 & \textbf{Memory allocation sizes can vary if sufficient for program needs} \\
& & Ground truth: \code{mov} \reg{r0}, \imm{\#400} \\
& & Predicted: \code{mov} \reg{r0}, \imm{\#800} \\
\midrule
P103 & 52 & \textbf{Different instruction sequences can achieve same logical result} \\
& & Ground truth: \code{cmp} \reg{r3}, \imm{\#0}; \code{and} \reg{r3}, \reg{r3}, \imm{\#1}; \code{rsblt} \reg{r3}, \reg{r3}, \imm{\#0} \\
& & Predicted: \code{rsbs} \reg{r2}, \reg{r3}, \imm{\#0}; \code{and} \reg{r3}, \reg{r3}, \imm{\#1}; \code{and} \reg{r2}, \reg{r2}, \imm{\#1}; \code{rsbpl} \reg{r3}, \reg{r2}, \imm{\#0} \\
\midrule
P69 & 50 & \textbf{Constants can be loaded directly or from literal pool} \\
& & Ground truth: \code{mvn} \reg{r3}, \imm{\#-2147483648} \\
& & Predicted:\newline \code{ldr} \reg{r3}, \imm{.L8}; \code{.L8:} \code{.word} \imm{2147483647} \\
\bottomrule
\end{tabular}}
\end{table}
\vspace{0.2cm}

\begin{table}[!h]
\small
\caption{Complex Variation Patterns with Multiple Differences}
\label{tab:complex_variations}
\resizebox{\linewidth}{!}{\begin{tabular}{@{}p{0.8cm}p{0.8cm}p{6.4cm}@{}}
\toprule
\textbf{Prog ID} & \textbf{Edit Dist} & \textbf{Combined Patterns and Examples} \\
\midrule
P128 & 78 & \textbf{Multiple Optimization Patterns:} \\
& & Groud truth: \code{mul} \reg{r1}, \reg{r2}, \reg{r3} \\
& & Predicted: \\
& & \code{lsl} \reg{r1}, \reg{r2}, \imm{\#2}; \\
& & \code{add} \reg{r1}, \reg{r1}, \reg{r2} \\
\midrule
P113 & 74 & \textbf{Memory and Instruction Patterns:} \\
& & Ground truth: \\
& & \code{str} \reg{r1}, \mem{[fp, \#-12]} \\
& & \code{mov} \reg{r3}, \reg{r2} \\
& & \code{add} \reg{r3}, \reg{r3}, \imm{\#4} \\
& & Predicted: \\
& & \code{str} \reg{r1}, \mem{[fp, \#-8]} \\
& & \code{add} \reg{r2}, \reg{r2}, \imm{\#4} \\

\bottomrule
\end{tabular}}
\end{table}

For incorrect implementations, we focused particularly on cases with small edit distances to understand how subtle differences can lead to functional failures (Table~\ref{tab:critical_errors}). These cases often involve critical errors in immediate values, register management, or memory access patterns that fundamentally alter the program's behavior.

\begin{table}[!h]
\large
\centering
\caption{Analysis of Critical Errors with Small Edit Distances}
\label{tab:critical_errors}
\resizebox{\linewidth}{!}{\begin{tabular}{@{}p{0.8cm}p{0.8cm}p{9.9cm}@{}}
\toprule
\textbf{Prog ID} & \textbf{Edit Dist} & \textbf{Example} \\
\midrule
P37 & 1 & \textbf{Incorrect immediate value causes wrong division factor and early loop termination} \\
& & Ground truth: \code{asr} \reg{r2}, \reg{r2}, \imm{\#2} \\
& & Predicted: \code{asr} \reg{r2}, \reg{r2}, \imm{\#1} \\
\midrule
P127 & 1 & \textbf{Array index offset error causes wrong element comparison} \\
& & Ground truth: \code{sub} \reg{r3}, \reg{r3}, \imm{\#2} \\
& & Predicted: \code{sub} \reg{r3}, \reg{r3}, \imm{\#1} \\
\midrule
P63 & 12 & \textbf{Register overwrite corrupts loop counter before multiplication} \\
& & Ground truth: \code{mov} \reg{r0}, \reg{r2}; \code{ldr} \reg{r1}, \mem{[r3, r1, lsl \#2]}; \code{mul} \reg{r0}, \reg{r0}, \reg{r1} \\
& & Predicted: \code{ldr} \reg{r0}, \mem{[r3, r1, lsl \#2]}; \code{mul} \reg{r0}, \reg{r0}, \reg{r1} \\
\midrule
P153 & 17 & \textbf{Incorrect instruction sequence fails to compute absolute value} \\
& & Ground truth: \code{sub} \reg{r2}, \reg{r2}, \reg{r3}; \code{cmp} \reg{r2}, \imm{\#0}; \code{rsblt} \reg{r2}, \reg{r2}, \imm{\#0} \\
& & Predicted: \code{sub} \reg{r1}, \reg{r2}, \reg{r3}; \code{eor} \reg{r2}, \reg{r1}, \reg{r2}; \code{sub} \reg{r2}, \reg{r2}, \reg{r1} \\
\midrule
P47 & 19 & \textbf{Mismatched memory access offsets cause incorrect data retrieval} \\
& & Ground truth: \code{str} \reg{r1}, \mem{[fp, \#-404]}; \code{ldr} \reg{r2}, \mem{[fp, \#-404]} \\
& & Predicted: \code{str} \reg{r1}, \mem{[fp, \#-404]}; \code{ldr} \reg{r2}, \mem{[r3, \#-20]} \\
\bottomrule
\end{tabular}}
\end{table}
\begin{figure*}[!t]
    \centering
    \small
    \includegraphics[width=\textwidth]{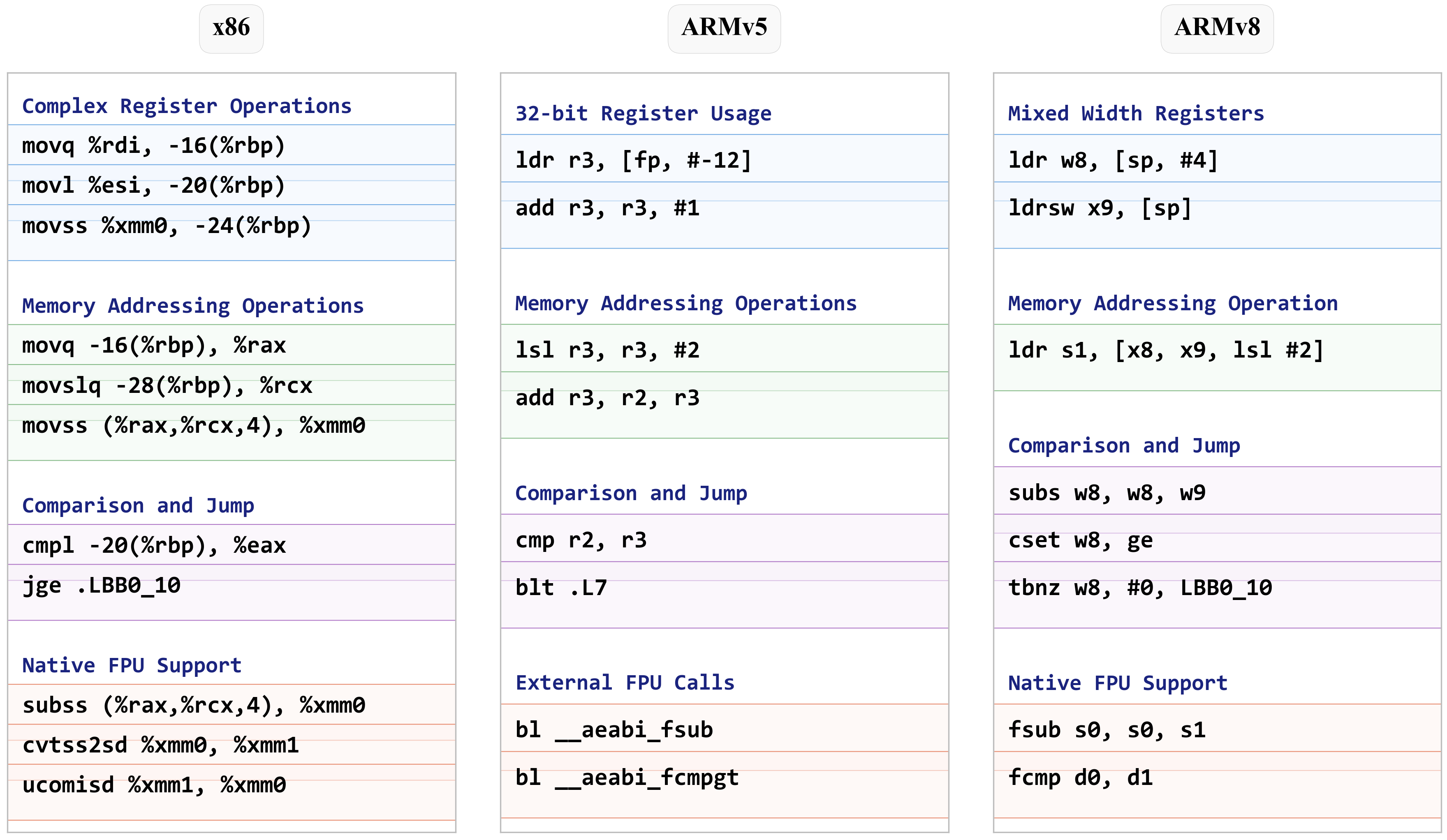}
    \caption{Example Assembly Code Generated from Identical Source Program for x86, ARMv5, and ARMv8, with aligned assembly segments highlighted by functionality.}
    \label{fig:assembly-comparison}
\end{figure*}

\subsection{Assembly Code Comparison Across Architectures}
\label{app-compare-archs}

Figure~\ref{fig:assembly-comparison} provides sample assembly code segments across different ISAs, all for the same program. We showcase some interesting cases, including how ARM has major differences between v5 and v8, despite both being RISC architectures. As the ISA evolves, the binary compatibility changes—a program compiled for ARMv5 might not be compatible with ARMv8.

Further, we highlight distinct challenges for transpilers in understanding certain constructs, such as mixed-width register operands, comparison-and-jump definitions, and the varying number of instructions needed to achieve the same functional objective. These samples showcase how ISAs are truly unique "languages," and why we believe (and evaluate) the efficacy of an LLM in transpiling between them.

\subsection{Alignment}
\label{appendix:assembly_comparison}

Building upon the \emph{Guess \& Sketch} method proposed by Lee et al.~\cite{lee2024guess} for cross-architecture transpilation using BART, we investigated the application of this neural machine translation model to convert x86 assembly code into ARM assembly code. However, BART's limitation of a 1024-token context window poses significant challenges when dealing with large assembly functions. This difficulty is further exacerbated by the substantial differences between x86 and ARM instruction sets, which necessitate the preservation of complex semantic relationships that often extend beyond the fixed window size.

To address these challenges, we attempted a semantic-aware code segmentation pipeline that facilitates effective translation while maintaining both local instruction patterns and the overall program structure across different architectures. Intuitively, we aimed to "elevate" the code up to the IR level to implement alignment.

The goal targeted precise code alignment within the limitations of BART's 1024-token context window. Our initial experiments with fixed-window segmentation resulted in a modest accuracy rate of 4.40\%, highlighting the inadequacy of simple segmentation methods in preserving essential program semantics. By implementing our semantic-aware pipeline, we significantly improved the translation accuracy to 16.46\%, effectively maintaining both local and global semantics during cross-architecture assembly translation. However, more work needs to be done, and we found that the larger context window size, as described in the main paper, effectively side-stepped this issue of alignment.

\end{document}